\documentclass{aa}
\usepackage[]{psfig,epsfig}
\usepackage[]{graphics,graphicx}
\newcommand{\be}{\begin{equation}}
\newcommand{\en}{\end{equation}}
\newcommand{\bear}{\begin{array}}
\newcommand{\enar}{\end{array}}

\newcommand{\etal}{{\it et al.}}

\newcommand{\simgt}{\lower.5ex\hbox{$\; \buildrel > \over \sim \;$}}
\title{Signatures of high energy protons in pulsar winds}
\author{Elena Amato\inst{1},Dafne Guetta\inst{1},Pasquale Blasi\inst{1}}
\authorrunning{Amato et al.}
\begin{document}
\institute{INAF/Osservatorio Astrofisico di Arcetri, Largo E. Fermi 5,I-50125 Firenze, Italy}
\offprints{E.Amato\\
e-mail:amato@arcetri.astro.it}
\date{Submitted------------- --,2002; accepted --------- --,2002}

%
%

\abstract{The resonant cyclotron absorption model (Hoshino \etal \cite{hosh2}, Gallant and Arons \cite{yves2}) is very successful in describing particle acceleration in plerions. A sensible prediction of this model is the presence of a substantial amount of relativistic protons in pulsar winds. Although difficult to detect, these protons may show up through their interactions either with the photons in the plerion environment or with the thermal gas in the supernova ejecta. Inelastic proton-proton (p-p) collisions are expected to be very effective in young objects, resulting in a copious production of neutral and charged pions. Charged pions produced during the first few hundred years after the supernova explosion may have time to decay into muons, whose subsequent decay may provide an additional source of electrons and positrons in these nebulae, that sums up to the pulsar input. These secondary leptons evolve just as the pairs in the pulsar wind, and signatures of their presence could be found, in principle, even in the synchrotron spectrum of older objects. $pp$ collisions may remain fairly efficient even in moderately old objects resulting in the production of TeV $\gamma$-rays and neutrinos. We apply our calculations to the case of the Crab Nebula, the best studied plerion insofar, and find that existing data already allow to infer interesting constraints on the physical properties of the Crab pulsar wind.}
\maketitle
%
%
\section{Introduction}
The presence in pulsar winds of a baryonic component coexisting with the
leptonic one is still an open question. 
According to the most successful model so far for particle acceleration
at the pulsar wind termination shock (Hoshino \etal \cite{hosh2},
Gallant and Arons \cite{yves2}), baryons should not only be present
in the wind, but even carry most of its energy. This makes it mandatory
to investigate all possible ways in which signatures of their presence can
be found. 

Plerions result from the interaction of a pulsar with its surrounding
medium. According to the standard picture, pulsars lose most of their
energy in the form of a magnetized relativistic wind, whose material
component is made mostly of electron-positron pairs. The wind starts
out cold and with a highly relativistic expansion velocity. At some
distance from the pulsar the wind then needs to be slowed down in order
to match the boundary conditions imposed by the surrounding supernova
ejecta expanding at subrelativistic speed. This happens at a termination
shock where most of the outflow bulk energy is randomized. Here is
where particle
acceleration presumably occurs, giving rise to the non-thermal particle
spectrum that is the source of the plerion synchrotron and Inverse
Compton radiation. 

According to both pulsar magnetospheric theories and the available data, 
the termination shock is most likely a relativistic transverse shock, 
meaning that the magnetic field is almost perpendicular to the flow 
velocity. Standard shock acceleration mechanisms ({\it i.e.} diffusive
shock acceleration or Fermi I process, and shock drift acceleration)
are generally believed to be ineffective at this kind of shocks
({\it e.g.} Hoshino \etal \cite{hosh2}).

An acceleration mechanism that is able to overcome the difficulties of
the former two is based on resonant cyclotron absorption (RCA) in a 
baryon-loaded plasma. Here we briefly review it in order to make clear
the assumptions we make on the proton spectrum and its relation with
the basic wind parameters.

Within the framework of the RCA model (Hoshino \etal \cite{hosh2}, 
Gallant \& Arons \cite{yves2}), the ions present in the wind
have a Larmor radius $r_{L_i}$ which is $A m_p/Z m_e$ times larger than that
of the pairs, being both species cold in the wind rest frame.
When the heavier particles meet the shock in the pairs, this
appears to them as an infinitesimally thin discontinuity in the
magnetic field. The sudden enhancement of the magnetic field is expected
to cause
them to start a bunched gyration, accompanied by the collective 
emission of cyclotron waves. The fundamental frequency of these waves
is the ion Larmor frequency, but also higher harmonics are excited
up to the pair gyration frequency. The latter can be resonantly absorbed
by the pairs and cause their acceleration, while the ions slowly
thermalize after crossing the shock. Efficient acceleration requires
the ions to be energetically dominant in the wind, although the pairs
are dominant by number.

After crossing the shock the ions drift towards the outer parts of the 
nebula with a velocity of order $c\ r_{L_i}/R_N$ and slowly thermalize 
to a temperature that is equivalent to a Lorentz factor $\Gamma$ of 
order of the initial wind Lorentz factor.
The full thermalization takes several gyration periods, so, depending
on the initial wind Lorentz factor it may never be achieved within
the nebula.

While drifting towards the edge of the nebula the ions may interact
with the plerion radiation field and with the thermal material of the
ejecta, which partly penetrates the synchrotron bubble due to 
Rayleigh-Taylor instabilities. Both these interactions would lead to
pion production. The products of the subsequent pion decays could lead to
observable signatures if the interaction rate were high enough.
The most direct signature would be the instantaneous emission of
high energy photons and neutrinos. 

The latter are potentially the
preferential channel to look at, since their detection would be
unequivocally associated with the presence of protons. In fact,
a few recent papers have considered the question of neutrino
production from protons accelerated in pulsar magnetospheres.
Bednarek \& Protheroe (\cite{bedprot}) computed the flux of high
energy neutrinos and $\gamma$-rays that would be expected 
from nuclear collisions of protons deriving from the decay of neutrons 
produced by the photodisintegration of accelerated nuclei in a
pulsar magnetosphere. Beall \& Bednarek (\cite{beall}) computed 
the neutrino flux that would be produced by high energy protons 
accelerated in the magnetosphere of a young fast rotating pulsar
surrounded by a supernova envelope,
and Bednarek (\cite{bednarek}) considered the effect of this contribution
on the overall neutrino background. 

None of these previous investigations explicitly considered the protons that
should be part of the pulsar wind if the RCA model for shock acceleration 
is actually at work in plerions. These are exactly the protons we
consider in this paper, with the aim of finding constraints from
observations on the characteristics of the pulsar wind.
In addition to computing the multi-TeV neutrinos and photon fluxes
we also worry about the secondary electrons
and positrons produced by $\pi^\pm$ decay. If protons are present
in fact, this process acts as an extra source of pairs in the plerion
which is expected to be especially active when the plerion is young
and compact, but could in principle leave relatively long-lasting
traces in the nebular radiation spectrum.

In the following we study the efficiency of the pion production
processes
as a function of time during the plerion evolution. We  
approximate the plerion as spherical and expanding at a constant
velocity. For the evolution of the nebular non-thermal emission
we use the Pacini \& Salvati (\cite{PS}) model (PS hereafter)
for a homogeneous plerion. 

%
%
\section{Pion production and energy losses}
\label{sec:s1}
Charged and neutral pions can be generated either by photo-pion production 
off the photons in the nebula or by inelastic p-p scattering against the
thermal gas in the ejecta. While the decay of neutral pions results in
$\gamma$-ray production, the decay of charged particles results in neutrinos
and electron-positron pairs (we will refer to both as electrons).

We describe the efficiency of both photomeson production and p-p 
collisions through a parameter, $f^\pi$, defined as the probability
for a single relativistic proton to create a pion before losing energy 
through other kinds of interactions:
\begin{equation}
\label{fpieq}
f^\pi={\rm min} \left[1, \frac{t_{\rm life}}{t^\pi} \right] \ ,
\end{equation}
where $t_{\rm life}$ is the proton life-time against losses different
from p-$\gamma$ and p-p, and $t^\pi$ is the timescale for the process
of pion production.
The luminosity output in pions will be:
\begin{equation}
\label{lpieq}
L_\pi=f^\pi L_p\ ,
\end{equation}
with $L_p$ the wind luminosity in protons.

In the presence of an intense magnetic or radiation field, charged pions
may lose a considerable fraction of their energy through synchrotron or
Inverse Compton emission before decaying into muons. The same could happen 
to muons before decaying into electrons and neutrinos. If any of these
effects were relevant, this would lead to a suppression 
of the neutrino and electron fluxes at high energies. We shall define the
quantities $f^\mu$ and $f^\nu$ as:
\begin{equation}
\label{fmufnueq}
f^\mu={\rm min} \left[1,{{t^\pi}_{\rm losses} \over 
{t^\pi}_{\rm decay}} \right]\, \, \, \, \,
\, \, \, \, \, \, \, \, \, \,
f^\nu={\rm min} \left[1, {{t^\mu}_{\rm losses} \over 
{t^\mu}_{\rm decay}} \right]\ ,
\end{equation}
where $t^\pi_{\rm decay}=2.6 \times 10^{-8}\ \gamma_\pi\ {\rm s}$ and
$t^\mu_{\rm decay}=2.2 \times 10^{-6}\ \gamma_\mu \ {\rm s}$.
Since the pion energy is roughly evenly shared between the final 
products of its decay, we can write 
the number of neutrinos, electrons and $\gamma$-rays injected in the
nebula per unit time and unit energy interval as:
\begin{eqnarray}
J_\nu(\epsilon_\nu) &\approx& 4\ \left(1+f^\nu[3\epsilon_\nu]\right)\ f^\mu[4\epsilon_\nu]\ \left. {dN_{\pi^\pm} \over dE_{\pi^\pm} dt} \right|_{E_{\pi^\pm}=4\epsilon_\nu}\ , 
\label{numdoteqnu}
\\
J_e(\epsilon_e)&\approx& 4\ f^\mu[4\epsilon_e]\ f^\nu[3\epsilon_e]\ \left. {dN_{\pi^\pm} \over dE_{\pi^\pm} dt} \right|_{\epsilon_{\pi^\pm}=4\epsilon_e}\ , 
\label{numdoteqel}
\\
J_\gamma(\epsilon_\gamma)&\approx&4\ \left. {dN_{\pi^0} \over dE_{\pi^0} dt} \right|_{E_{\pi^0}=2\epsilon_\gamma} \ .
\label{numdoteqgam}
\end{eqnarray}
An important point to be noticed in the above set of equations is that in
writing Eq.(\ref{numdoteqnu}) we take into account only muon neutrinos,
since these are the particles that neutrino telescopes are able to detect. 
The possible effects of neutrino oscillations on the results of our
calculations will be discussed in section \ref{sec:s4}.


%
%
\subsection{Photo-meson production}
\label{sec:s101}
In the case of photo-meson production the target for high energy protons
is the plerion emission. 
The fractional energy loss rate of a proton with energy 
$E_p$ (= $ \Gamma\, m_p\, c^2$) due to pion production results in
(Waxman \& Bahcall, 
\cite{waxbah}):
\begin{eqnarray}
\label{tpgesteq}
t_{p \gamma}^{-1}(E_{p})&\simeq&\frac{2^{p+1}}{p+2}\ 
\sigma_{\rm peak}\ \xi_{\rm peak}\ \frac{\Delta \epsilon}
{\epsilon_{\rm peak}} \left(\frac{d}{{\rm R}_N} \right)^2\ 
\frac{ F_\nu(\nu_{\rm p})}{4\ \pi\ h} \nonumber \\
&\simeq&
2.5 \times 10^{-16} \frac{2^{p+1}}{p+2} 
\left( \frac{d_{\rm kpc}} {R_{\rm pc}} \right)^2
F_{\nu_{\rm p}}[{\rm mJy}]\ {\rm yr}^{-1}.
\end{eqnarray}
Here we have treated the plerion as homogeneous and used the fact that 
the photon spectrum is a power law,
$F_\nu \propto \nu^{-p}$. We have also made the approximation that the
main contribution to pion production comes from 
photon energies $\epsilon_\gamma \approx \epsilon_{\rm peak}$=0.3 GeV, 
where the p-$\gamma$ cross section peaks due to the $\Delta$ resonance.
The numerical values are obtained using: $\sigma_{\rm peak}\ =\ 5 \times 10^{-28}\ {\rm cm}^2$, 
$\xi_{\rm peak}$ =0.2, $\Delta \epsilon$=0.2 GeV,
$\nu_{\rm p}=\epsilon_{\rm peak}/ (\Gamma\ h)$ and $\beta_p \simeq 1$.

%
%
\subsection{Inelastic nuclear collisions}
\label{sec:s102}
The energy loss-rate of a relativistic proton due to inelastic nuclear
collisions can be estimated as
\begin{equation}
\label{tppeq1}
t_{pp}^{-1} \approx \zeta\ n_t\ \sigma_0\ c\ ,
\end{equation}  
where $n_t$ is the target density, $\sigma_0=5 \times 10^{-26} {\rm cm}^2$
and $\zeta\simeq 20 \%$ is the average fraction of energy lost by the proton.

As we mentioned in the introduction, the estimate of the effective target
density, $n_t$, is the main uncertainty in our calculation.
The distribution of the thermal material, in all cases when it is observed, 
is actually far from uniform. The matter of the 
ejecta penetrates the synchrotron nebulae due to Rayleigh-Taylor 
instabilities,
giving rise to concentrations in the form of ``filaments''. The total
amount of material contained in these filaments is not easy to estimate 
with confidence. In fact the possibility that the gas detected through 
thermal emission hides some colder and denser concentrations cannot be
excluded. A comparably difficult task is to estimate, from the projection 
on the plane of the sky, the fraction of the nebular volume that these 
filaments occupy, their so called ``filling factor''.

Moreover the effective target density for relativistic protons, which is
our reason for worrying about the distribution of thermal matter in plerions, 
is also affected by the intensity and structure of the magnetic field in and 
around the ``filaments'', which are not known. The magnetic field affects 
in fact the physics of proton propagation and could in principle be an 
obstacle to the penetration of protons in the thermal gas, leading to a 
decrease of the value of $n_t$, or conversely be the cause of proton 
trapping in the filaments, resulting in the opposite effect. 
  
We shall summarize all these unknowns in a single parameter
$\mu$, which we also take to be time-constant. We define the effective
density of material in terms of the average gas density in the plerion as:
\begin{equation}
\label{denseq}
n_t=\mu\ \bar{n}_t=\mu\ \frac{M_{\rm N}}{m_p}\ \frac{3}{4 \pi {\rm R}_N^3}\ =\
10\ \mu\ \frac{M_{{\rm N} \odot}}{{\rm R}_{\rm pc}^3}\ {\rm cm}^{-3}\ ,
\end{equation}
where $M_N$ is the mass of the ejecta contained in the filaments, which
we also take to be time-constant.
Using Eq. (\ref{denseq}) in Eq. (\ref{tppeq1}) we obtain:
 \begin{equation}
\label{tppesteq}
t_{pp}^{-1} \approx 10^{-7}\ \mu\ \frac{M_{{\rm N} \odot}}{{\rm R}_{\rm pc}^3}\ {
\rm yr}^{-1}\ ,
\end{equation}
which, when compared with Eq. (\ref{tpgesteq}) indicates nuclear collisions 
as the main mechanism for pion production in plerions, if $\mu$ is not very
much less than 1.

%
%
\subsection{Energy losses}
\label{sec:s103}
The energy distribution of particles after injection is determined, in
a plerion, mainly by the following loss-mechanisms: escape and expansion
losses, and radiative cooling through synchrotron and Inverse Compton
emission. We approximate the plerion as a homogeneous spherical bubble
filled with a uniform magnetic field and expanding at constant velocity
$v$, and estimate the time-scales of the different processes under these 
assumptions.
 
The rate of adiabatic cooling is simply:
\begin{equation}
\label{texpeq}
t_{\rm exp}^{-1}= \left(\frac {v} {{\rm R}_N}\right) \simeq 10^{-3} \frac{{\rm V}_3}{{\rm R}_{\rm pc}}\ {\rm yr}^{-1}\, 
\end{equation}
where ${\rm R}_{\rm pc}$ is the nebular radius in {\it parsecs} and ${\rm V}_3$
is the plerion expansion velocity in units of $10^3$ km/s.

The rate of escape of a particle of mass $m_s$ and Lorentz factor $\Gamma^s$,
whose Larmor radius in the nebular magnetic field is $r_L$, can be estimated, 
based on $\vec \nabla B$ drift, as:
\begin{equation}
\label{tesceq}
{t^s_{\rm esc}}^{-1}= \frac{c\ {\rm r}_L}{{\rm R}_N^2}\simeq 3 \times 10^{-3}\ \frac {\Gamma^s_6} {{{\rm B}_{-4}}\ {\rm R}_{\rm pc}^2}\ \frac{m_s}{m_p}\ {\rm yr}^{-1},
\end{equation}
where $B_{-4}$ is the nebular magnetic field in units of $10^{-4}$ G and 
$\Gamma^s_6$ is the particle Lorentz factor in units of $10^6$.

The rate of synchrotron cooling is:
\begin{eqnarray}
\label{tsynceq}
{t^s_{\rm sync}}^{-1} &=& {4 \over 3}\ \sigma_T\ c\ \left({m_e \over m_s}\right)^2\ {\Gamma^s \over m_s c^2}\ U_B \nonumber \\
 & \simeq & 7 \times 10^{-14}\ \Gamma^s_{6}\ B_{-4}^2  \left( {m_p \over m_s} \right)^3 {\rm yr}^{-1}\ .
\end{eqnarray}

Finally, the cooling rate due to Inverse Compton Scattering (ICS hereafter) is
\begin{eqnarray}
\label{ticseq}
{t^s_{\rm ICS}}^{-1}&=& { 4 \over 3} \sigma_T c\ \left({m_e \over m_s} \right)^2 
{\Gamma_s \over m_s c^2} \left\{U_{\rm cmb}+{{\rm R}_N \over c} \right. \nonumber \\
& \times  & \left[ 
\int_{\nu_{\rm min}}^{\nu_{\rm KN}} S_\nu d\nu + \left. 
{9 \over 32}  \nu_{\rm KN}^2 \int_{\nu_{\rm KN}}^{\nu_{\rm max}} 
{S_\nu \over \nu^2} d\nu \right] \right\}\  ,
\end{eqnarray}
where $\nu_{\rm KN}=m_sc^2/\Gamma^s/h$ is the transition frequency between
interaction in the Thompson and Klein-Nishima regimes, $U_{\rm cmb}$ is the
energy density of the cosmic microwave-background photons, and $S_\nu$ is 
the volume emissivity of the plerion.

%
%
\section{Proton signatures during the plerion history}

\subsection{Plerion evolution}
\label{sec:s2}
All the rates in the previous section depend on quantities that change
with time during the plerion evolution. We determine the dependency on 
time of the different quantities within the framework of the PS model.
We consider the spin-down power of the pulsar to be continuously 
converted into injection of fresh particles and magnetic field energy.
The spin-down luminosity of the pulsar can be parametrized as:
\begin{equation}
\label{spindowneq}
L(t)={L_0 \over (1+t/\tau)^\alpha}\ ,
\end{equation}
where the characteristic spin-down time $\tau$ is a few hundred years,
and the index $\alpha$ is expected to be around 2 if the breaking is
purely electromagnetic.

Treating the plerion as homogeneous and expanding at constant velocity,
the magnetic field evolution, resulting from the competition between the
continuous pulsar supply and adiabatic losses, can be determined analytically.
For times shorter or comparable with $\tau$, one can make the approximation:
\begin{equation}
\label{bevoleq}
B(t)=\sqrt{3 L_B \over v^3 t^2}\ ,
\end{equation}
where $L_B$ is the fraction of the pulsar luminosity that goes into
magnetic energy.

The other quantity we need to know as a function of time in order to
estimate ICS losses and the rate of photo-meson production is the 
plerion volume emissivity $S_\nu$, due to non-thermal radiation.
To this purpose, we need to determine the time-dependences of the
spectral density $N(E,t)$ of the emitting particles (pairs). 
For the wind particles we assume that they are injected with an initial
spectrum in the form of a broken power-law: 
\begin{equation}
\label{partinjeq}
J_w(E)=K_w \left\{
\matrix{
\left(E /E_1 \right)^{-\gamma_1}\ ; \, \, \, \, E_{\rm min} \leq E \leq  E_1\cr
 \left(E/E_1\right)^{-\gamma_2}\ ; \, \, \, \,  E_1 \leq  E \leq E_{\rm max} \  \cr
}
\right. \ ,
\end{equation}
where $K_w$ is related to the fraction of pulsar luminosity that is converted 
into pairs, $L_{\rm pairs}$ through:
\begin{eqnarray}
\label{Keq}
K_w(t)&=&{L_{\rm pairs}(t) \over E_1^2} \left\{ 
{1 \over 2-\gamma_1} \left[1-\left({E_{\rm min} \over E_1}\right)^{2-\gamma_1}\right] \right. \nonumber \\
&+&\left.
{1 \over 2-\gamma_2} \left[{\left(E_{\rm max} \over E_1\right)}^{2-\gamma_2}-1\right]
\right\}^{-1} \ .
\end{eqnarray}
In the following we shall assume $E_1$ to be constant with time 
and independent on the wind Lorentz factor: we take $E_1=1.5\ {\rm TeV}$,
roughly corresponding to the energy of the particles radiating at the
frequency at which the high energy 
break of the spectrum of the Crab Nebula is observed.  
As to $E_{\rm max}$, this will depend on both time and the wind Lorentz
factor, due to the dependence on the latter of the proton energy and
on the dependence on the former of the nebular magnetic field strength. 
In fact we assume the maximum
energy to which the pairs can be accelerated to be given by 
$E_{\rm max}={\rm min}(E_p,E_{\rm cut})$, with $E_{\rm cut}$
determined by the condition 
$T_{\rm acc}(E_{\rm cut}) = T_{\rm losses}(E_{\rm cut})$, {\em i.e.}
the acceleration time cannot exceed the time-scale for energy losses.
Assuming $T_{\rm acc} \approx \Omega_L^{-1}$ (with $\Omega_L$ the particle
Larmor frequency) and the losses as due
to synchrotron emission in the average nebular magnetic field, one
finds $E_{\rm cut}(t) = m_e c^2 \sqrt{6 \pi e/(\sigma_T B(t))}$. 

We mentioned that charged pion decay provides an additional source of
pairs in the nebula. Their evolution is completely analogous to that of
the wind particles, so that apart from the different initial spectrum,
the treatment below applies to both components. 

The spectral distribution at a time $t$, $N(E,t)$, is found as a function 
of the injection spectrum $J=J_w+J_e$ (see Eq. (\ref{numdoteqel})), as: 
\begin{equation}
\label{njeq}
N(E,t)=\int_E^{E_{\rm max}}J(E_i,t_i)\ {\partial  t_i \over \partial E}(E,E_i,t)\
dE_i\ ,
\end{equation}
where $t_i$ is the time at which the particle was born and $E_i$ its
initial energy. $N(E,t)$ is readily computed 
once $\partial t_i / \partial E$ is known as a function of $E$ and $t$.

The needed relation between $E$, $E_i$ and $t_i$ may be found from the 
equation describing the
single particle energy evolution due to synchrotron and adiabatic losses: 
\begin{equation}
\label{partevol}
{1 \over E\ t} -{1 \over E_i\ t_i}=\int_{t_i}^t c_1 {B^2(s) \over s} ds\ ,
\end{equation}
where $c_1=(1/6\pi)\ \sigma_T\ c/(m_e c^2)^2$ is 
the constant relating the synchrotron power to magnetic field strength
and particle energy.
For $\partial t_i / \partial E$ we find:
\begin{equation}
\label{dtieq}
{\partial t_i \over \partial E}={E_i\ t_i^2(E,t) \over E^2\ t}\
\left[ 1 + {2\ E_i \over E_b(t_i)} \right]^{-1}\ ,
\end{equation}
where we have used the definition of the so called break energy:
\begin{equation}
\label{ebdefeq}
E_b(t)={2 \over c_1\ t\  B^2(t)}\ .
\end{equation}

For times shorter than $\tau$, when $B(t) \propto t^{-1}$ (Eq.
(\ref{bevoleq})), so that $E_b(t)\propto t$, one finds:
\begin{eqnarray}
\label{tieq}
t_i&=&{t \over 2 (1 +E/E_b(t))} {E \over E_i} \nonumber \\
& \times & \left\{ 1+
\left[1+4{E_i^2 \over E E_b(t)}
\left(1+ {E \over E_b(t) } \right) \right]^{1/2} \right\}\ ,
\end{eqnarray}
and the solution of Eq. (\ref{njeq}) can be approximated as:
\begin{equation}
\label{nappreq}
N(E,t) \approx {2\ t \over \gamma_1+\gamma_2} \left\{
\matrix{
J(E,t_0), \ , \, \, \, \, E \leq  E_b(t)\cr
(E_b(t)/E) J(E,t_0) \ , \, \, \, \, E_b(t) \leq  E\cr
}
\right.
\end{equation}
with $t_0\simeq 0$ for $t<\tau$ and $t_0 \simeq \tau$ afterwards.

\begin{figure}[h!!!]
\centerline{\includegraphics*[bb = 50 355 575 690, height=5.7cm]{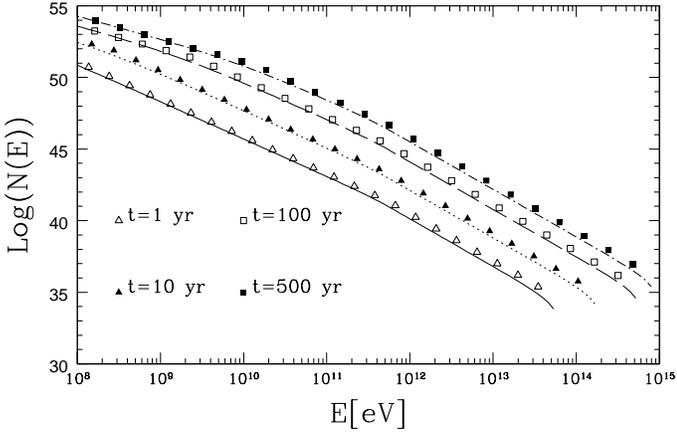}}
\caption{\footnotesize{The time evolution of the wind particle spectrum 
in the
case $\gamma_1=1.5$, $\gamma_2=2.2$, 
$L_{\rm pairs}=L_B=10^{39}\ {\rm erg}\ {\rm s}^{-1}$. The different curves refer 
to different epochs after the supernova explosion: $t=1 {\rm yr}$ (solid),
$t=10 {\rm yr}$ (dotted), $t=100 {\rm yr}$ (dashed), 
$t=500 {\rm yr}$ (dot-dashed). The points trace, at each time, 
the approximation given in the text.}}
\label{Nfig}
\end{figure}

The evolution of the wind particle spectrum as a function of time is
plotted in Fig. (\ref{Nfig}) for the case $\Gamma=10^6$, $\gamma_1=1.5$,
$\gamma_2=2.2$ and $L_0=3 \times 10^{39}\ {\rm erg}\ {\rm s}^{-1}$,
with $L_B=L_{\rm pairs}=L_p=(1/3) L$. 
We plot the full solution of the problem taking into account the
real time-dependences arising from the varying pulsar input (Eq. 
(\ref{spindowneq}) with $\tau=700$ yr) together
with the approximate expression in Eq. (\ref{nappreq}).  

Once $N(E,t)$ and $B(t)$ are known, the volume emissivity $S_\nu$ that 
enters Eq. (\ref{ticseq}) is computed from synchrotron theory as:
\begin{equation}
\label{snusynceq}
S^{\rm sync}_\nu(\nu,t)={c_1 B \over 2 c_2} {3 \over 4\ \pi\ R_N^3} 
\sqrt{{\nu \over c_2\ B}}\ N\left[\sqrt{{\nu \over c_2\ B} }\ , t \right]\ ,
\end{equation}
where $c_2=0.3\ e\ c/(2\ \pi\ (m_e\ c^2)^3)$. Knowing $S^{\rm sync}_\nu$
one can also compute the ICS emissivity of the plerion, $S^{\rm ICS}_\nu$,
following,
for instance, Blumenthal \& Gould (\cite{blum}). 
From the total emissivity $S_\nu=S^{\rm sync}_\nu+S^{\rm ICS}_\nu$ 
we also obtain
the flux $F_\nu$ that appears in Eq. (\ref{tpgesteq}): 
\begin{equation}
\label{fnueq}
F_\nu(\nu)={R_N^3 \over 3\ d^2}\ S_\nu(\nu)\ .
\end{equation}

%
%
\subsection{Pion production efficiency}
\label{sec:s3}
The time-evolution of the magnetic and radiation field in the nebula
as derived in the previous section was used to compute the time-history
of pion production efficiency. 

In the plots below we show the behaviour of $f^\mu$, $f^\nu$, 
$f^\pi_{p \gamma}$ and $f^\pi_{pp}$ as a function of time after the 
supernova explosion for various values of the pulsar luminosity $L$, of the 
wind Lorentz factor $\Gamma$ and of the remnant expansion velocity $v$. 
In all the plots in this section we assumed the pulsar energy input 
to be evenly shared between magnetic field, protons and pairs: 
$L_B=L_p=L_{\rm pairs}=(1/3)\ L_0$.
To compute the radiation field, we assumed the injection spectrum of the
pairs as in Eq. (\ref{partinjeq}) with $\gamma_1=1.5$ and $\gamma_2=2.2$.
We consider a monoenergetic distribution of protons corresponding to the
Lorentz factor of the wind.

\begin{figure}[h!!!!]
\centerline{\includegraphics*[bb = 18 155 565 690,height=8cm]{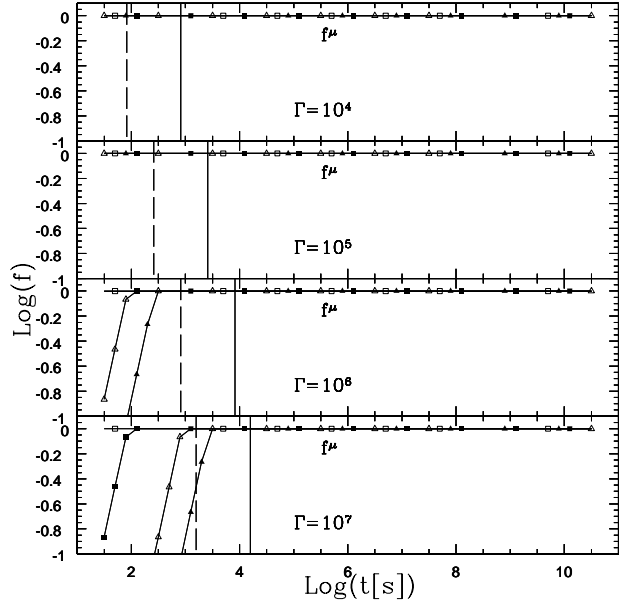}}
\caption{\footnotesize{The time evolution of $f^\mu$ for different values of 
the pulsar luminosity, of the remnant expansion velocity and of the wind 
Lorentz factor. The curves plotted in different panels are computed for 
different values of the wind Lorentz factor.
The different curves in each panel are marked as explained below.
Empty triangles: 
$L[{\rm erg}/{\rm s}]=3 \times 10^{39}$, $v[{\rm km}/{\rm s}]=10^3$.
Empty squares:
$L[{\rm erg}/{\rm s}] =3 \times 10^{39}$,$v[{\rm km}/{\rm s}]=10^4$.
Filled triangles:
$L[{\rm erg}/{\rm s}] =3 \times 10^{40}$,$v[{\rm km}/{\rm s}]=10^3$.
Filled squares:
$L[{\rm erg}/{\rm s}] =3 \times 10^{40}$,$v[{\rm km}/{\rm s}]=10^4$.
The vertical lines represent the time at which the optical depth for neutrinos
drops below 1 in the case when $v=10^3\ {\rm km}/{\rm s}$ (solid) and 
when $v=10^4\ {\rm km}/{\rm s}$ (dashed).}}
\label{fmufig}
\end{figure}

In Fig. (\ref{fmufig}) we plot $f^\mu(E_\pi,t)$ for the value of 
$E_\pi$ corresponding to the highest energy pions produced for each of the
values of $\Gamma$ considered. 
\begin{figure}[h!!!!]
\centerline{\includegraphics*[bb = 18 155 565 690,height=8cm]{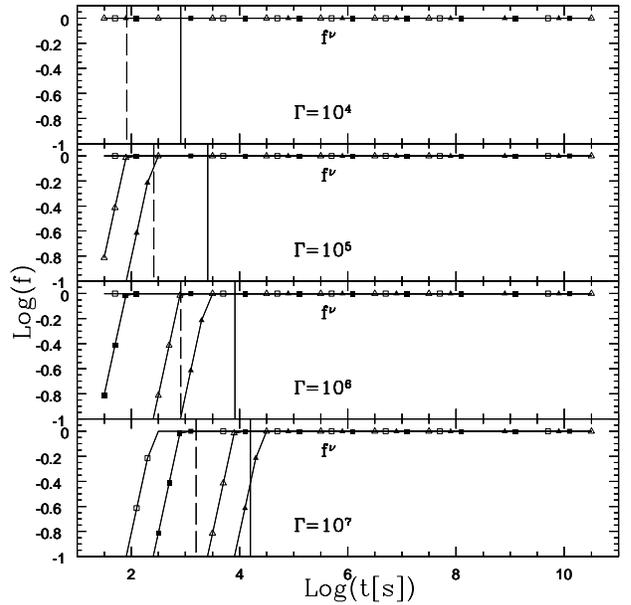}}
\caption{\footnotesize{The time evolution of $f^\nu$ for the highest energy
pions corresponding to different values of 
the pulsar luminosity, of the remnant expansion velocity and of the wind 
Lorentz factor. Same notation as in Fig. (\ref{fmufig}).}}
\label{fnufig}
\end{figure}

It is apparent from the figure that for Lorentz factors up to $10^5$
pions never suffer any severe energy losses, not even at early times,
when the magnetic and radiation field in the plerion are most intense.
For energetic enough pions, instead, radiation losses at early times do 
become important, decreasing the decay probability by a factor that can 
be of order $10^2$ for the lowest considered value of the expansion velocity 
and highest considered pulsar luminosity (filled triangles).

Losses at early times are found to be due mainly to synchrotron emission 
and their dependence on $v$ and $L$ arises from the relation between the 
magnetic field intensity and these two quantities (Eq. (\ref{bevoleq})).
It is interesting to notice that in all the cases considered, $f^\mu$ 
becomes 1 at very early times, and actually earlier than the time at which 
the remnant stops being optically thick to neutrinos, which we mark as 
a reference time in Fig. (\ref{fmufig}). This reference time is computed
considering the fact that the main source of optical depth for high energy 
neutrinos are nuclear collisions that give rise to muons: 
$\nu + X \rightarrow \mu + Y$. The cross section for this process 
({\em e.g.} Learned \& Mannheim \cite{learned}) is a function of the 
neutrino energy alone at the energies we consider.
Hence, the time at which $\tau_\nu=R_N\ n_t\ \sigma_{\nu X}$ drops 
below unity for the highest energy neutrinos corresponding to each
value of the Lorentz factor we consider depends only on the 
Lorentz factor itself and on the density evolution of the remnant
({\em i.e.} on its expansion velocity, once the total mass is fixed
and assumed as uniformily distributed). One has:
$\tau_\nu \propto \Gamma^\beta/v$, with $\beta=0.5$ if 
$\Gamma<4 \times 10^5$ and $\beta=0.2$ otherwise.  

The same {\it legenda} as for Fig. (\ref{fmufig}) applies to Fig. 
(\ref{fnufig}), where the evolution of $f^\nu$ is shown (again in the case
when it is minimum, {\it i.e.} for the maximum muon energies). 

As we could expect, radiation losses for muons are more severe than for 
pions,
and for the most energetic we are considering (lower right panel) $f^\nu$
actually becomes 1 later than $f^\mu$.

It is important to notice that as far as only the non-thermal radiation 
field is taken into account, synchrotron losses are always dominant with
respect to losses due to ICS, according to our
calculation. The latter can
become important, though, if also the thermal radiation field is
taken into account, but this effect should be limited to earlier times
than that at which $\tau_\nu$ drops below 1 (Beall \& Bednarek \cite{beall}).

\begin{figure}[h!!!!]
\centerline{\includegraphics*[bb = 18 155 565 690,height=8cm]{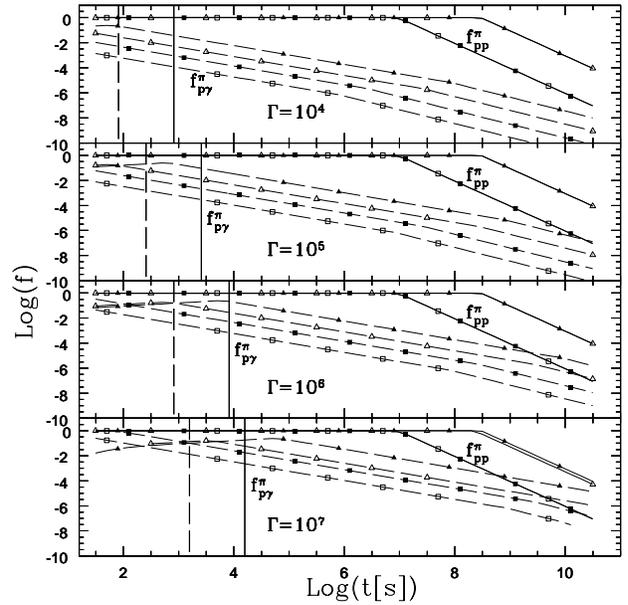}}
\caption{\footnotesize{The time evolution of $f^\pi$ for different values of 
the pulsar luminosity, of the remnant expansion velocity and of the wind 
Lorentz factor. The solid curves refer to the efficiency of nuclear collisions
while the dashed ones show the efficiency of photo-meson production. Curves
referring to different values of the pulsar luminosity and remnant expansion
velocity are marked according to the same notation as for Figs. (\ref{fmufig})
and (\ref{fnufig}).}}
\label{fpifig}
\end{figure}

As far as $f^\pi$ is concerned, we computed the efficiency of both
photo-meson production and p-p collisions. For the latter process
we considered $1\ M_\odot$ of material uniformily distributed within the
remnant and used $\mu=1$ in the calculations. The results are shown
in Fig. (\ref{fpifig}).

We find that radiation losses are never effective to 
limit the efficiency of p-p collisions, not even in the very intense 
initial magnetic field of a bright, slowly expanding plerion.
Conversely, for high enough values of the wind Lorentz 
factor $\Gamma$ and low enough values of the wind expansion velocity, 
synchrotron losses are the main limit to photo-meson production. 
However, we can see from Fig. \ref{fpifig} that photomeson production 
never reaches 
efficiencies higher than about $30 \%$, at least for the prototypical 
spectrum that has been assumed here.

It should be noticed, finally, that
in the plots shown above for $f^\pi$, $f^\mu$ and $f^\nu$, when computing
both ICS losses and photo-meson production we have only
included the photon spectrum deriving from synchrotron and ICS emission of
the wind particles. In doing that we have neglected the thermal emission 
from the gas and dust in the filaments and the radiation deriving from 
pion decay. The consistency of the latter approximation was verified 
{\it a posteriori}: after calculating the radiation flux from the
products of pion decay we checked that adding this contribution to
the target radiation for ICS losses did not change the results plotted
in Figs. (\ref{fmufig})-(\ref{fpifig}).

\subsection{Photon and neutrino fluxes from a ``young Crab''}
\label{sec:s4}
In this section we use the theory introduced above to compute both the
electromagnetic and neutrino emission from a young plerion with Crab-like
parameters, as a function of time after the supernova explosion.
We consider a pulsar whose initial spin-down power is 
$L_0= 3 \times 10^{39}$ erg/s. 
We assume, as in section \ref{sec:s3}, that $L_0$ is evenly converted 
into electron, proton and magnetic field energy: $L_B=L_p=L_{\rm pairs}=(1/3) 
L_0$. The distance of the pulsar and the plerion expansion velocity 
are taken to be the same as for the Crab, and 1 $M_\odot$ of thermal material
uniformily distributed in the plerion volume is considered as the target
for p-p collisions. 

\begin{figure}[h!!!!]
\centerline{\includegraphics*[bb = 18 153 568 690,height=8cm]{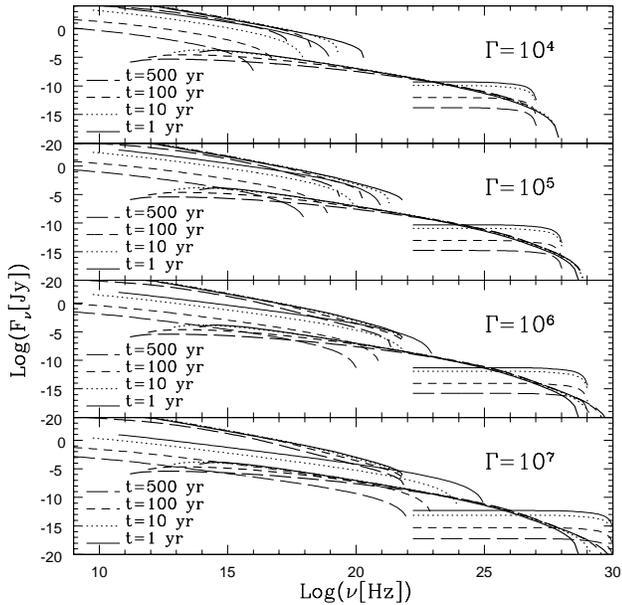}}
\caption{\footnotesize{Electromagnetic spectrum of a ``young Crab''
at different times after the supernova explosion. The thick curves 
correspond to the synchrotron (at low energies) and ICS (at higher energies)
emission of the wind pairs, while the lighter curves refer to
the emission of the pion decay products. The different panels correspond
to different values of the wind Lorentz factor $\Gamma$ and the different
line types refer to different times.}}
\label{youngspec}
\end{figure}

The time evolution of the pion production efficiency in such a plerion
was computed in section \ref{sec:s3} (curves marked with empty squares in
Figs. (\ref{fmufig})-(\ref{fpifig})).
Taking the results from that section we estimate the global photon spectrum
of the plerion. The high energy part is entirely due to $\pi^0$ decay, while 
at low energies we estimate the synchrotron emission of both wind and
secondary pairs together with the ICS emission of the former. The overall 
spectrum at different times during the
plerion evolution is reported in Fig. (\ref{youngspec}).

It is apparent from the figure that, for the set of parameters used,
the radiation due to the secondary pairs could only be detectable over the
emission due to the wind particles, at early times (up to few tens of 
years after the supernova explosion) and at photon energies in the
range 100 MeV-1 TeV.  

We notice that $\gamma$-rays from $\pi^0$ decay should be 
detected even by present TeV detectors
during the first few tens of years after the supernova explosion,
if a similar plerion was born within
the galaxy. The $\pi^0$ decay $\gamma$-ray flux always exceeds the ICS
emission at high enough energies for times upto several tens of years
at least. 

\begin{figure}[h!!!!]
\centerline{\includegraphics*[bb = 18 155 565 690,height=8cm]{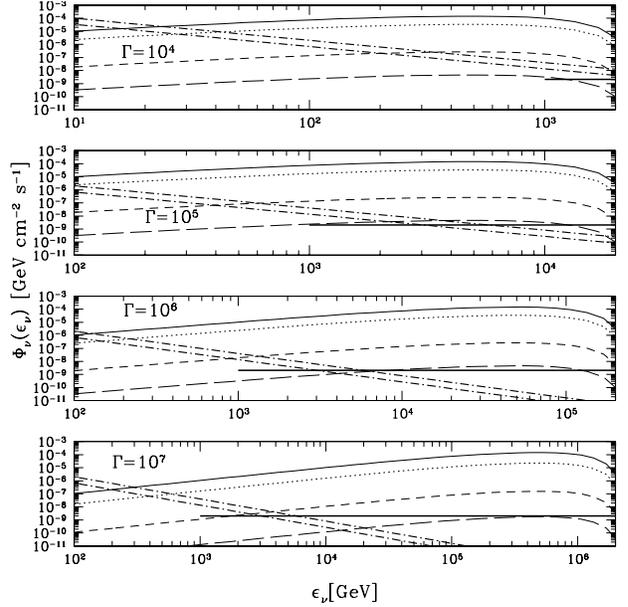}}
\caption{\footnotesize{Neutrino spectrum of a ``young Crab''. The notation is
the same as for Fig. (\ref{youngspec}). The solid thick line is the detection
threshold for ICE-cube (Hill \cite{hill}), while the dot-dashed line is the
atmospheric neutrino background in a $2^\circ \times 2^\circ$ bin.}}
\label{youngnu}
\end{figure}

In Fig. (\ref{youngnu}) we show the expected neutrino flux from the 
same plerion. We find that this is generally well above the atmospheric
background of a detector with an angular resolution of 
$2^\circ \times 2^\circ$. Moreover
a galactic object described by our set of parameters could give a flux 
of neutrinos already detectable by telescopes like AMANDA II for the 
first ten years. It is clear from the figure that the probability of
detecting such objects will increase considerably with the upcoming 
${\rm km}^2$ neutrino telescopes, since their lower threshold allows
neutrino detection even from older plerions, if the pulsar wind
contains a non-negligible fraction of protons.

%
%
\section{Constraining the parameters of the Crab pulsar wind}
\label{sec:s5}
In this section, we are going to use the Crab Nebula, the prototype
plerion, to discuss how the information coming from high
energy photon and neutrino detection could be used to constrain the 
physics of plerions. 

The parameters playing a role in determining the flux of high energy
photons and neutrinos are all quite well known for the Crab Nebula, 
except for the parameters that enter in the determination of $\mu$, 
the wind Lorentz factor $\Gamma$, and the 
fraction of the pulsar spin-down power that goes into acceleration
of protons. The relevant parameters for our purposes are, from the
equations in section \ref{sec:s1}: 
the plerion age $t =950 {\rm yr}$,
its distance $d \simeq 2\ {\rm kpc}$,
its expansion velocity, $V_{\rm exp} = 1.5 \times 10^3\ {\rm km}\ {\rm s}^{-1}$,
the average magnetic field intensity $B \simeq 3 \times 10^{-4}$ G,
and finally the amount of mass contained in the filamentary shell.
For the latter, the most recent estimates give $M_N=(4.6\pm 1.8) M_\odot$
(Fesen \etal \cite{fesenetal}). We shall consider a value $M_N=1 M_\odot$
and values of $\mu$ in the range $1<\mu<20$, according to the upper
limit on $n_t$ derived by Atoyan \& Aharonian (\cite{at&ara}), based
on the compatibility between the bremsstrahlung $\gamma$-rays and the
observed flux from the Crab Nebula above 1 GeV. We notice that if
the mass estimated by Fesen \etal (\cite{fesenetal}), were uniformily
distributed in the nebula, this would correspond to $\mu=5$.

Inserting the numbers just quoted into the equations in section
\ref{sec:s1} we can estimate all the relevant time-scales.
One can readily see that what is important  
for computing $f^\pi$ is just the 
comparison between $t_{\rm esc}$ and $t_{\rm exp}$ (whose relative
importance depends on $\Gamma$) and $t_{pp}$, since
the other three time-scales are much longer than these. From
Eqs. (\ref{fpieq}), (\ref{tppesteq}), (\ref{texpeq}) and (\ref{tesceq}),
we have:
\begin{eqnarray}
f^\pi&=&{\rm min} \left[1, {t_{\rm exp} \over t_{pp}} {\rm min}
\left(1,{t_{\rm esc} \over t_{\rm exp}} \right) \right] \nonumber \\
&=&{\rm min} \left[ 1, 2.8 \times 10^{-5} \mu\ {\rm min} 
\left(1, {2.2 \over \Gamma_6} \right) \right],
\label{fpiesteq}
\end{eqnarray}
where $\Gamma_6$ is the wind Lorentz factor in units of $10^6$.

For the same values of the parameters the muon losses due to synchroton
emission or ICS (Eqs. (\ref{tsynceq}) and (\ref{ticseq})) are completely 
negligible on the time-scale of their decay, even for the highest 
energy we consider. The same is true for pions, whose 
proper decay times are about a hundred times shorter.

\begin{figure}[ht!!!]
\centerline{\includegraphics*[bb=20 140 570 700 , height=8cm]{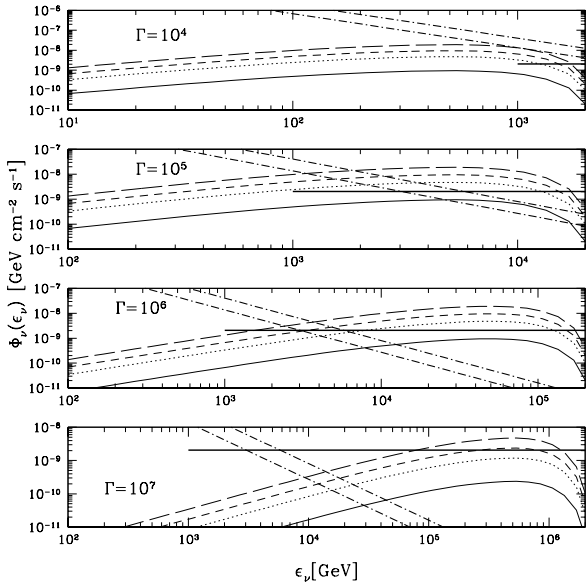}}
\caption{\footnotesize{The flux of neutrinos expected from
the Crab Nebula as a function of energy, for different values of 
the parameter $\mu$ and for different values of the wind Lorentz factor 
$\Gamma$. The different curves in each panel correspond to $\mu=1$ (solid),
$\mu=5$ (dotted), $\mu=10$ (dashed) and $\mu=20$ (long-dashed). 
The considered
values of $\Gamma$ range from $10^4$ to $10^7$, as specified in each panel.
The solid thick line is the detection threshold for IceCube 
(Hill \cite{hill}), while the dot-dashed line is the atmospheric
neutrino background in a $2^\circ \times 2^\circ$ bin.
We assume 60 \%
of the pulsar wind energy to be carried by protons.}}
\label{CrabNu}
\end{figure}

\begin{figure}[ht!!!]
\centerline{\includegraphics*[bb=20 140 570 700 , height=8cm]{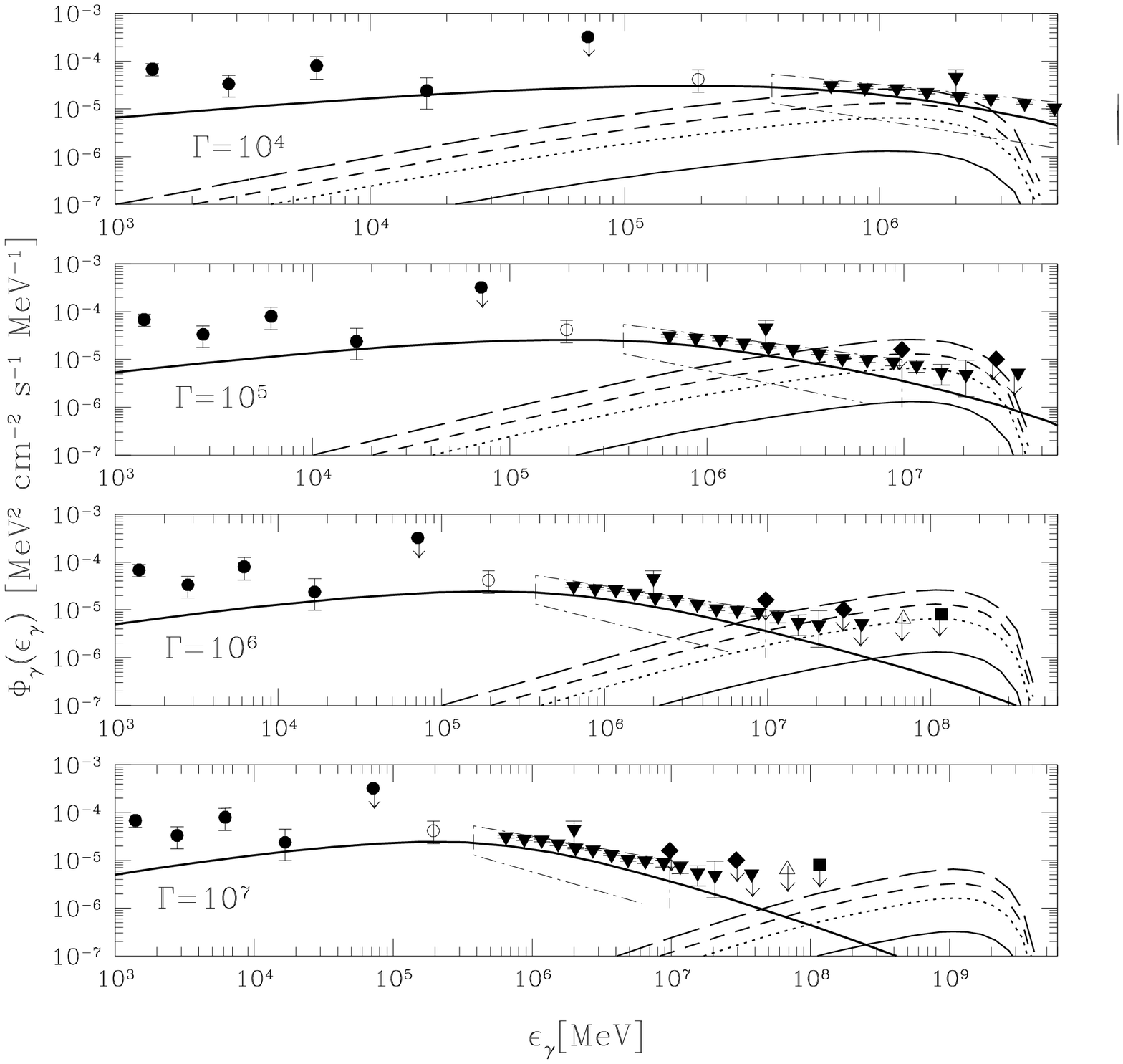}}
\caption{\footnotesize{The flux of high energy photons expected from the 
Crab Nebula, for different values of the parameter $\mu$ and
of the wind Lorentz factor $\Gamma$. The notation is the same as in Fig.
(\ref{CrabNu}). The expected
fluxes are compared with the available data and upper limits
(de Jager \etal \cite{dej96},
Aharonian \etal \cite{aharet} and references therein). Also shown as a thick 
curve in each panel is the ICS flux computed based on this model.}}
\label{CrabGam}
\end{figure}

In Figs. (\ref{CrabNu}) and (\ref{CrabGam}), we plot, respectively, the 
expected neutrino and photon fluxes from the Crab Nebula as a function 
of energy, for different assumptions on the wind Lorentz factor and 
different values of the parameter $\mu$. 


We have assumed that protons are 
the main energy carriers in the wind so that we put in them about $60 \%$
of the pulsar spin-down luminosity 
$L_p = 0.6\ L_{\rm Tot} \simeq 3 \times 10^{38}\ {\rm erg}\ {\rm s}^{-1}$,
being $L_{\rm Tot}=5 \times 10^{38}\ {\rm erg}\ {\rm s}^{-1}$. 
Our choice of the energy fraction into baryons comes from the fact that
this is the percentage of the pulsar power which is required to 
be in baryons if the luminosity contrast of the Crab Nebula wisps is to
be explained through magnetic field compression due to the ions gyration
in the shocked background pair plasma (Spitkovsky \& Arons \cite{anatoly}). 

The neutrino and photon fluxes were computed from Eqs. 
(\ref{numdoteqnu}) and (\ref{numdoteqgam}) respectively,
with the pion differential spectrum per unit time
computed in the scaling approximation for the p-p scattering cross-section
(Blasi \cite{blasi}). In the scaling regime a monoenergetic proton distribution
with energy $E_p$ leads to the injection of pions described as:
\begin{equation}
{dN_\pi \over dE_\pi dt}=K_\pi {E_\pi}^{-1}\ g_\pi(E_\pi/E_p)\ ,
\label{dnpieq}
\end{equation} 
where the function $g_\pi$ is
\begin{equation}
g_\pi(x)=(1-x)^{3.5}+e^{-18x}/1.34\ ,
\label{gpieq}
\end{equation}  
and $K_\pi$ is found from the condition
\begin{equation}
f_\pi L_p = \int_{m_\pi c^2}^{E_p} {dN_\pi \over dE_\pi dt} E_\pi dE_\pi\ ,
\label{kpidefeq}
\end{equation}
leading to:
\begin{equation}
K_\pi=f^\pi {L_p \over E_p} \left( \int_0^1 g_\pi(x) dx \right)^{-1}\ .
\label{kpieq}
\end{equation}

It is apparent from Fig. (\ref{CrabNu}) that, for the generally favoured value
of $\Gamma \simeq 10^6$ (Kennel \& Coroniti \cite{kc1} 
and  \cite{kc2}; Gallant \& Arons \cite{yves2}), and for the value of 
$L_p \simeq 0.6 L_{\rm tot}$ favoured by Gallant \& Arons (\cite{yves2})
and Spitkovsky \& Arons (\cite{anatoly}),
the flux of neutrinos from the Crab Nebula would be well above the
atmospheric neutrino background already for $\mu=1$.
The chances of detecting multi-TeV neutrinos from the Crab Nebula with
IceCube look promising, based on this figure.

However, part of the section of the parameter space that would lead to 
a detectable flux of neutrinos can already be excluded based on the
available high energy $\gamma$-ray data. In Fig.(\ref{CrabGam}) we
plot the flux of high energy photons expected from $\pi^0$ decay.
It is clear that in some cases the computed fluxes at multi-TeV energies
are above the measurements and upper limits. The latter, on the other
hand, can be comfortably explained, whithin the error bars, as due to
ICS (de Jager \etal \cite{dej96}; Atoyan \& Aharonian \cite{at&ara};
Aharonian \etal \cite{aharet}). Even within the framework of our crude 
model, which not only treats the plerion as homogeneous but also
neglects the contribution to the target photon field of the ambient 
thermal emission, the computed ICS flux is only slightly below the
TeV data points.    

In the following we shall review our expectation for the flux of neutrinos 
from the Crab Nebula in the light of the constraints arising from 
$\gamma$-ray observations. 

Before discussing these limits, however, we should also consider the
other possibility we mentioned to put constraints on the three
unknowns $\mu$, $L_p$ and $\Gamma$. This is offered, at least in principle,
by the signatures that may be left by the secondary pairs in the
synchrotron radiation spectrum of the nebula.
These pairs would be injected in the Nebula according to Eq.
(\ref{numdoteqel}), but the injection rate (described by Eq. (\ref{dnpieq})
and depending on $f^\pi$ and $L_p$ through Eq. (\ref{kpieq}))
has to be considered now as a function of time. Since we are mainly 
interested in low energy pairs, with long synchrotron life-times, we
need to take into account the whole history of pion production in the
nebula. This is done as discussed in sections \ref{sec:s2} and \ref{sec:s3}.

\begin{figure}[h!!!]
\centerline{
\includegraphics*[bb= 20 145 570 700,height=8cm]{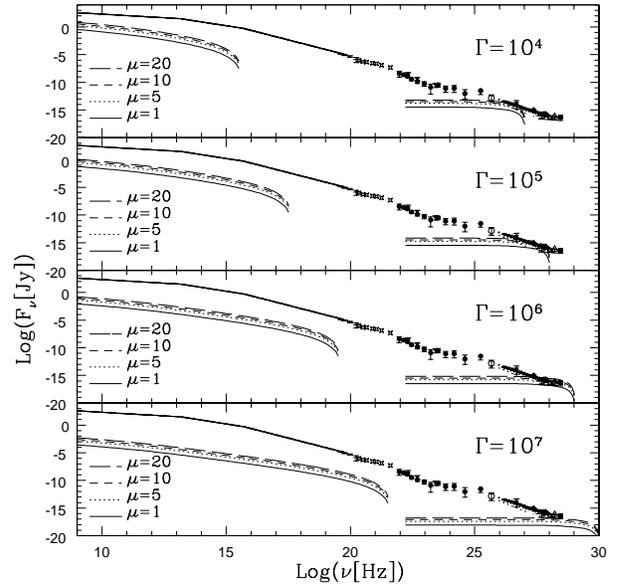}}
\caption{\footnotesize{Consequences of pion production on the electromagnetic
spectrum of the Crab Nebula. The thick curve and the points represent the
observed emission, while the thin curves are fluxes from our model.
At low frequencies the model flux is due to
synchrotron emission associated to secondary electrons and positrons, while
the contribution at high frequencies is the same shown in Fig. (\ref{CrabGam})
and comes from $\pi^0$ decay. The different panels refer to the emission computed
for different values of the wind Lorentz factor while the different line-types 
are associated to different values of the parameter $\mu$ as specified in the 
each panel.}}
\label{CrabSync}
\end{figure}

We show in Fig. (\ref{CrabSync}) the comparison between the Crab 
Nebula observed radiation 
spectrum and the synchrotron flux from secondary pairs computed in the 
case when $L_p(t)=0.6\ L_{\rm Tot} (t)$, with $L_{\rm Tot} (t)$ described 
by Eq.
(\ref{spindowneq}), where $\tau=730\ {\rm yr}$ and $\alpha=2.3$ 
(Groth \cite{groth}). In the same figure, we also show the contribution
arising from $\pi^0$ decay, in the $\gamma$-ray part of the spectrum.
It is apparent that the emission produced by the secondary pairs,
according to our model, is a completely negligible fraction of the 
present day synchrotron emission of the Crab Nebula and cannot provide
any additional constraint on the parameters we are interested in.

\begin{figure}[h!!!]
\centerline{
\includegraphics*[bb= 20 145 570 540,height=5cm]{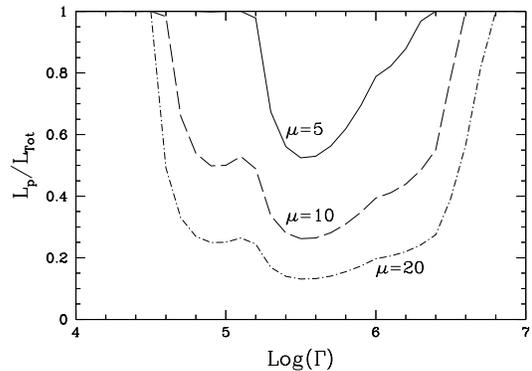}}
\caption{\footnotesize{The allowed parameter space for 
$\Gamma$ and $L_p/L_{\rm Tot}$ as constrained through the comparison
with observations of the photon fluxes from pion decay computed within
the framework of our model.
The allowed regions for each of the considered value of $\mu$ are the
ones lying under the curve of corresponding type: $\mu=5 \rightarrow$ 
solid,
$\mu=10 \rightarrow$ dashed, $\mu=20 \rightarrow$ dot-dashed. No constraints
can be put on the wind parameters for $\mu=1$.}}
\label{LpMax}
\end{figure}

In Fig. (\ref{LpMax}) we use the information that can be extracted 
from high energy data (Fig. (\ref{CrabGam})) to restrict the allowed 
parameter space for
the values of $L_p/L_{\rm Tot}$, $\Gamma$ and $\mu$.
The allowed regions, for each choice of $\mu$ are those lying below 
the corresponding curve, as explained in the figure caption.

It is interesting to notice that, if $\mu=10$ or larger, the possibility
that 60\% of the pulsar wind energy might be in protons with a
Lorentz factor $\Gamma=10^6$ would be excluded by the data.

\begin{figure}[h!!!]
\centerline{
\includegraphics*[bb= 20 170 570 505,height=5cm]{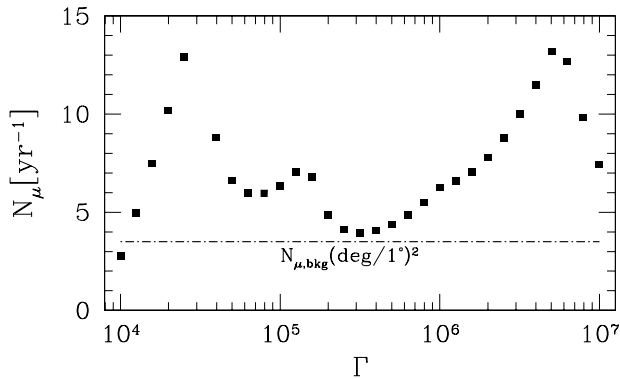}}
\caption{\footnotesize{The maximum number of muons per year that would be 
produced by neutrinos coming from the Crab Nebula in a ${\rm km}^2$ detector.
Also shown, as a dot-dashed line, are the background counts, assuming an angular 
resolution of $(1^\circ)^2$, such as IceCube should have.}}
\label{NumNu}
\end{figure}

We now use the constraints in Fig. (\ref{LpMax}) to estimate the
maximum number of neutrino events that one could expect to detect from 
the Crab Nebula in 1 yr of observation with a detector of area $1\ {\rm km}^2$,
such as IceCube will be.
The result of this calculation is shown in Fig. (\ref{NumNu}) where we 
plot the maximum number of muons per year as a function of the wind
Lorentz factor. $N_\mu$ is computed taking into account the probability
for a neutrino to interact with the matter in the detector and create
a muon. For each value of $\Gamma$ we consider the maximum allowed value
of $f^\pi L_p$ with the requirement that $\mu \le 20$ and 
$L_p \le 0.9 L_{\rm Tot}$. This determines $J_\nu(\epsilon_\nu)$.
$N_\mu$ is then found as:
\begin{equation}
 N_\mu={1 {\rm yr} \times  1 {\rm km}^2 \over 4\ \pi\ {\rm d}^2} \int_{1 {\rm TeV}}^{\epsilon_{\rm Max}} J_\nu(\epsilon_\nu) P_{\nu \mu}(\epsilon_\nu) d \epsilon_\nu
\label{numnueq}
\end{equation}
where the maximum neutrino energy is $\epsilon_{\rm Max}=m_p c^2 \Gamma/4$ and
$P_{\nu \mu}=1.3 \times 10^{-6} \epsilon_{\nu,{\rm TeV}}^\beta$ with $\beta=1$ if
$1\ {\rm TeV}<\epsilon_\nu<100\ {\rm TeV}$, and $\beta =0.5$ otherwise 
(Gaisser \etal \cite{pnumu}).

The background counts we show in Fig. (\ref{NumNu}) are those computed
for IceCube by Alvarez-Mu\~niz and Halzen (\cite{alvarez}) at the position 
of the Crab Nebula. The maximum number of neutrino events allowed by the
$\gamma$-ray data is above the background for basically the entire range
of values of $\Gamma$ we considered. The effect of neutrino oscillations 
would be to reduce these fluxes. For the case of maximal mixing, equal 
abundancies of neutrinos of the three flavors are obtained, so that the 
flux of muon neutrinos is about half of that calculated above. 

\section{Conclusions}
\label{sec:sf}
In this paper we have anylized the observational consequences that the
presence in pulsar winds of a proton component, energetically dominant,
might have. 

We have found that when the plerion is very young (up to 
a few tens of years after the supernova explosion) the main loss 
mechanism for these protons is likely to be pion production due to 
nuclear collisions. A large flux of neutrinos, $\gamma$-rays and 
secondary pairs can be expected during these early stages. The 
synchrotron emission 
from secondaries is likely to be completely negligible, while
the TeV photon and neutrino 
emission from a Galactic object could be detectable even by present-
day telescopes during the first few tens of years after the supernova 
explosion. Detection of neutrinos from older plerions should
await the upcoming ${\rm km}^2$ detectors.

We have also estimated the present-day TeV photon and neutrino
emission from the Crab Nebula. 
The existing TeV observations already exclude part of the parameter
space, so that not all combinations of $\Gamma$, $L_p$ and $\mu$ are
allowed.

Finally we have computed the maximum number of neutrino events allowed by the 
$\gamma$-ray data according to our modeling and found it to be larger than
the atmospheric neutrino background (as already suggested by Arons, 
\cite{arelba}). Such a flux is likely to
be detected by a ${\rm km}^2$ detector such as IceCube. Neutrino detection
can be therefore considered as a promising means to help clinch the
question of how large a fraction of the pulsar spin-down energy goes into
relativistic baryons.

\section*{Akcnowledgements}
We thank F. Pacini and M. Salvati for fruitful discussions and we are
grateful to J. Arons for reading and commenting an early 
version of the paper.
Finally, we thank the referee very much for his valuable suggestions. 
This work has been partly supported by the Italian Ministry of University
and Research (MIUR) under the grant Cofin-2001-02-10.


\begin{thebibliography}{}
%
\bibitem[2000]{aharet}
Aharonian F.A. \etal, 2000, ApJ, {\bf 539}, 317
%
\bibitem[2001]{alvarez}
Alvarez-Mu\~niz J., \& Halzen F., 2001, astro-ph/0.05408
%
\bibitem[1998]{arelba}
Arons J., 1998, Mem. S.A.It., {\bf 69}, 989
%
\bibitem[1996]{at&ara} 
Atoyan A.M. \& Aharonian F.A., 1996, MNRAS, {\bf 278}, 525
%
\bibitem[1997]{bedprot}
Bednarek W. \& Protheroe R.J., 1997, PRL, {\bf 79}, 2616)
%
\bibitem[2001]{bednarek}
Bednarek W., 2001, A\&A, {\bf 378}, L49
%
\bibitem[2002]{beall}
Beall J.H. \& Bednarek W., 2002, ApJ, {\bf 569}, 343
%
\bibitem[1990]{berebook}
Berezinskij V.S., Bulanov S.V., Dogiel V.A., Ginzburg V.L., Ptuskin V.S., 1990, ``Astrophysics of Cosmic Rays'' (North-Holland: Amsterdam)
%
\bibitem[2001]{blasi}
Blasi P., 2001, APh, {\bf 15}, 223
%
\bibitem[1970]{blum}
Blumenthal G.S. \& Gould R.J., 1970, Rev. Mod. Phys., {\bf 42}, 237
%
\bibitem[1997]{fesenetal}
Fesen R.A., Shull J.M., Hurford A.P., 1997, AJ, {\bf 113}, 354
%
\bibitem[1995]{pnumu}
Gaisser T.K., Halzen F. \& Stanev T. 1995, Phys. Rep., {\bf 258}, 173
%
\bibitem[1994]{yves2} 
Gallant Y. A. \& Arons J., 1994, ApJ, {\bf 435}, 230
%
\bibitem[1975]{groth} 
Groth E. J., 1975, ApJS., {\bf 29}, 431
%
\bibitem[2001]{NEMO}
Halzen F., 2001, in ``Intl. Symp. on High Energy Gamma Ray Astronomy'', Heidelberg, June 2000, astro-ph/0103195
%
\bibitem[2001]{hill}
Hill C.G., 2001, Proceedings of the ${\rm XXXVI}^{\rm th}$ Rencontres de Moriond,
astro-ph/0106064
%
\bibitem[1992]{hosh2}
Hoshino M. \etal, 1992, ApJ, {\bf 390}, 454
%
\bibitem[1996]{dej96} 
de Jager O.C. \etal, 1996, ApJ, {\bf 457}, 253
%
\bibitem[1984a]{kc1} 
Kennel C.F. \& Coroniti F.V., 1984a, ApJ, {\bf 283}, 694
%
\bibitem[1984b]{kc2} 
Kennel C.F. \& Coroniti F.V., 1984b, ApJ, {\bf 283}, 710
%
\bibitem[2000]{learned}
Learned J.G. \& Mannheim k., 2000, Ann. Rev. Nucl. Part. Sci., {\bf 50}, 679
%
\bibitem[1973]{PS}
Pacini F. \& Salvati M., 1973, ApJ, {\bf 186}, 249
%
\bibitem[1999]{anatoly} 
Spitkovsky A. \& Arons J., 1999, AAS, {\bf 195}, 3304
%
\bibitem[1997]{waxbah}
Waxman E. \& Bahcall J.N., 1997, Phys. Rev. Lett., {\bf 78}, 2292
%
\end{thebibliography}
\end{document}